\documentclass[aps,prd,preprint,preprintnumbers,showpacs]{revtex4-1}
\usepackage{graphicx}
\usepackage{amsmath}
\usepackage{caption}
\usepackage[section] {placeins}
\usepackage{slashed}
\usepackage[dvips]{color}
\usepackage{soul}
\usepackage{multirow}
\begin{document}

\title{Predicting the existence of  a 2.9 GeV $Df_0(980)$ molecular state }

\author{A.~Mart\'inez~Torres\footnote{amartine@if.usp.br}}
\author{K.~P.~Khemchandani\footnote{kanchan@if.usp.br}}
\author{ M.~Nielsen\footnote{mnielsen@if.usp.br} }
\author{ F.~S.~Navarra\footnote{navarra@if.usp.br}}
\preprint{}

 \affiliation{
Instituto de F\'isica, Universidade de S\~ao Paulo, C.P. 66318, 05389-970 S\~ao 
Paulo, SP, Brazil
}

\date{\today}

\begin{abstract}
A $D$-like meson resonance with mass around 2.9 GeV has been found in the 
$DK\bar K$ system using two independent  and different model calculations based
 on: (1) 
QCD sum rules and (2) solution of Fadeev equations with input interactions 
obtained from effective field theories built by considering both chiral and 
heavy quark symmetries. The QCD sum rules have been used to study the 
$D_{s^*0}(2317) \bar{K}$ and $D f_0(980)$ molecular currents. A resonance of mass  
2.926 GeV  is found with  the $D f_0(980)$ current. Although a state in the 
$D_{s^*0}(2317) \bar{K}$ current is also obtained, with mass around 2.9 GeV, the 
coupling of this state is found to be two times weaker than the one formed 
in $D f_0(980)$. On the other hand, few-body equations are solved for the 
$D K \bar{K}$ system and its coupled channels with the input $t$-matrices 
obtained by solving Bethe-Salpeter equations for the $D K$, $D\bar{K}$ and 
$K \bar{K}$ subsystems.  In this study a $D$-like meson with mass 2.890 GeV and full 
width $\sim$ 55 MeV is found to get dynamically generated when $D K \bar{K}$ 
gets reorganized as $D f_0(980)$. However, no clear signal appears for the 
$D_{s^*0}(2317) \bar{K}$ configuration. The striking similarity between the results 
obtained in the two different models indicates strongly towards the existence 
of a $D f_0 (980)$ molecule with mass nearly 2.9 GeV. 
\end{abstract}

\maketitle
 
\section{Introduction}
In the past years, the development of high energy facilities has lead to the 
discovery of a number of open and hidden charm resonances by collaborations 
like BABAR, Belle and BES~\cite{kroko,aubert,abe,choi,li} which, in turn, 
has motivated many  theoretical studies to understand the properties and nature 
of heavy flavor hadrons. Among which some of the heavily discussed states are 
$D_{s^{*}0}(2317)$, $X(3872)$, $Z^{+}(4430)$, whose properties have been 
studied within different models assuming different configurations like 
diquarks, tetraquarks, hybrids, hadron molecules, etc. (for a review see 
Refs.~\cite{mfrev,Swanson:2006st,Zhu:2007wz,Brambilla:2010cs}).

The understanding of the nature of the different mesons and baryons of the 
hadron spectra, in general, is a long standing puzzle in theoretical nuclear 
physics. QCD is the accepted fundamental theory describing the strong 
interactions in terms of the quarks and gluons which constitute the hadronic 
matter. However, while at high energies the theory becomes perturbative and has 
been successfully tested by the experiment, the situation is very different at 
low energies, where due to the confinement of the quarks the theory is not 
anymore perturbative and nonperturbative methods are needed to extract 
information about the properties of the hadrons.

To face this challenging issue different techniques have been developed. One of 
them is Lattice QCD, which in the last few years  has emerged as an important
 tool to extract information about hadronic observables like mass, phase shifts,
 etc. However, due to the large number of degrees of freedom present in QCD 
(quarks and gluons of different flavors and colors) numerical calculations 
involving big number of lattice points and small lattice spacing are very time 
consuming for natural values of the mass of the quarks. Although a lot of 
progress has been done in this area, there are still  some problems when 
addressing excited states which have decay 
channels~\cite{nakahara,mathur,basak,bulava,morningstar}.

Another alternative to study hadrons within the spirit of QCD is the method of 
QCD sum rules (QCDSR) (see Refs.~\cite{svz,rry,SNB,colangelo2} for a pedagogical information on this topic). 
In this formalism the hadrons are 
described in terms of their interpolating quark currents, with which a 
correlation function is built. One begins evaluating this correlation function 
at short distances, where the quark-gluon dynamics is essentially perturbative, 
and then nonperturbative corrections are added to it. This method has been 
widely used to understand the mass, coupling, decay width, etc., of many hadron 
states.

Yet another way to elucidate the nature and properties of mesons and baryons is 
based on the use of effective field theories built by taking into account 
unitarity, chiral symmetry and its spontaneous breaking. In this case, the 
hadrons are the degrees of freedom of the theory instead of the quarks which 
constitute them. In the last 20 years, there has been lot of activity in this 
field and many resonances have been found to have important meson-meson or 
meson-baryon components in their wave functions. Some of the states most widely 
discussed are the $\Lambda(1405)$, generated as a consequence of the interaction
 of the coupled channel system $\bar K N$ and 
$\pi\Sigma$~\cite{kaiser,osetramos,ollerulf,jido3,jido2}, and the $f_0(980)$ 
resonance, formed in the $K\bar K$ and $\pi\pi$ system
\cite{oller,pelaez,ollerprog}. Recently, this theory has been generalized to 
study the properties of hadronic systems in a finite volume and its value in 
the determination of related physical observables using the energy levels 
obtained in the finite volume and, thus, as a prospective tool for Lattice QCD 
calculations, has been 
shown~\cite{michael1,albaladejo,roca,martineztorres1,martineztorres2}.\\
\indent The above mentioned methods are in continuos development since the experimental 
access to higher and higher energies is becoming plausible and, consequently, 
more and more new states with heavy quarks are being found. Present time is 
thus ideal to study heavy hadron physics since model predictions can be 
immediately tested, which eventually helps in understanding the structure of 
hadrons. With this idea we present a study of the $DK\bar K$ system, which we 
find particularly interesting since the $DK$ and $K\bar K$ interactions are 
attractive in nature. In this manuscript, we have studied this system using two 
methods: QCDSR and Few-body equations. In the former case, we 
investigate the $D f_{0}(980)$ and $D_{s^*0}(2317) \bar K$ configurations, while 
in the latter we solve the Faddeev equations for the three-hadron system, where 
$f_{0}(980)$ and $D_{s^*0}(2317)$ are dynamically generated in the corresponding 
subsystems. As we shall see, we find a resonance with similar characteristics 
in both models. \\
\indent In the following, we first discuss the calculations based on QCD sum 
rules and the results found in it. Subsequently, we tackle with the formalism 
to solve the Faddeev equations and discuss the results obtained with it. Finally
 we draw some conclusions.
\section{ QCD sum rules}
We start our study based on the QCDSR by writing the interpolating 
molecular currents for the $D f_0$ and $D_{s^*0} \bar{K}$ systems as
\begin{eqnarray}
j^{D f_0} &=& i \left(\bar q_a \gamma_5 c_a\right) \left(\bar s_b s_b\right) \label{jdf0}\\
j^{D_{s^*0} \bar{K}} &=& i \left(\bar s_a  c_a\right) \left(\bar q_b \gamma_5 s_b\right),\label{jdsk}
\end{eqnarray}
where $a$ and $b$ are color indices, and $q$ represents a light quark ($u$ or $d$). Using these currents, we write the  two-point correlation function
\begin{equation}
\Pi (q^2) = i \int d^4x e^{i q\cdot x} \langle 0 \mid T \left[ j (x) j^{\dagger}(0)\right] \mid 0 \rangle, \label{corfn}
\end{equation}
which can be written in terms of the quark propagators by contracting all the 
quark anti-quark pairs (for more details see, for example, Ref~\cite{mfrev}).

This function is of a dual nature: it represents a  quark-antiquark fluctuation 
at short distances (or large negative $q^2$) and can be treated in perturbative
 QCD, while at large 
distances it can be related to hadronic observables.
The sum rule calculations are based on the assumption that in some range of 
$q^2$ both descriptions are equivalent. One, thus, proceeds by calculating 
Eq.~(\ref{corfn}) 
for both cases and by eventually equating them to obtain  information on the
 properties of the hadrons.

From the QCD side, for large momentum transfers,  Eq.~(\ref{corfn}) can be 
calculated, in the first approximation, by assuming the involved propagators as 
those of free quarks. However,
since we are finally interested in studying  the properties of hadrons,  the 
relevant energies are lower, where the distance between the quarks gets longer
and quark-gluon interactions, quark-antiquark pair creation becomes important. 
It is, thus, required to include the effect of the presence
of the gluons and quarks in the QCD vacuum.
For practical calculations, then, one resorts to
 the Wilson operator product expansion (OPE) method, where the correlation 
function is expanded in a series of local operators
\begin{equation}
\Pi^{\rm OPE} =\sum\limits_{n} C_n (Q^2) O_n \label{ope}.
\end{equation}
In Eq.~(\ref{ope}) the set $\{O^n\}$ contains all local gauge invariant operators 
expressible in terms of the gluon fields and the fields of light quarks and the 
coefficients $C_n(Q^2)(Q^2=-q^2)$, by construction, include only the 
short-distance domain and can, therefore, be evaluated perturbatively. 
Nonperturbative long-distance effects are contained only in the local operators. 

\begin{figure}[ht!]
\includegraphics[width= 0.95\textwidth]{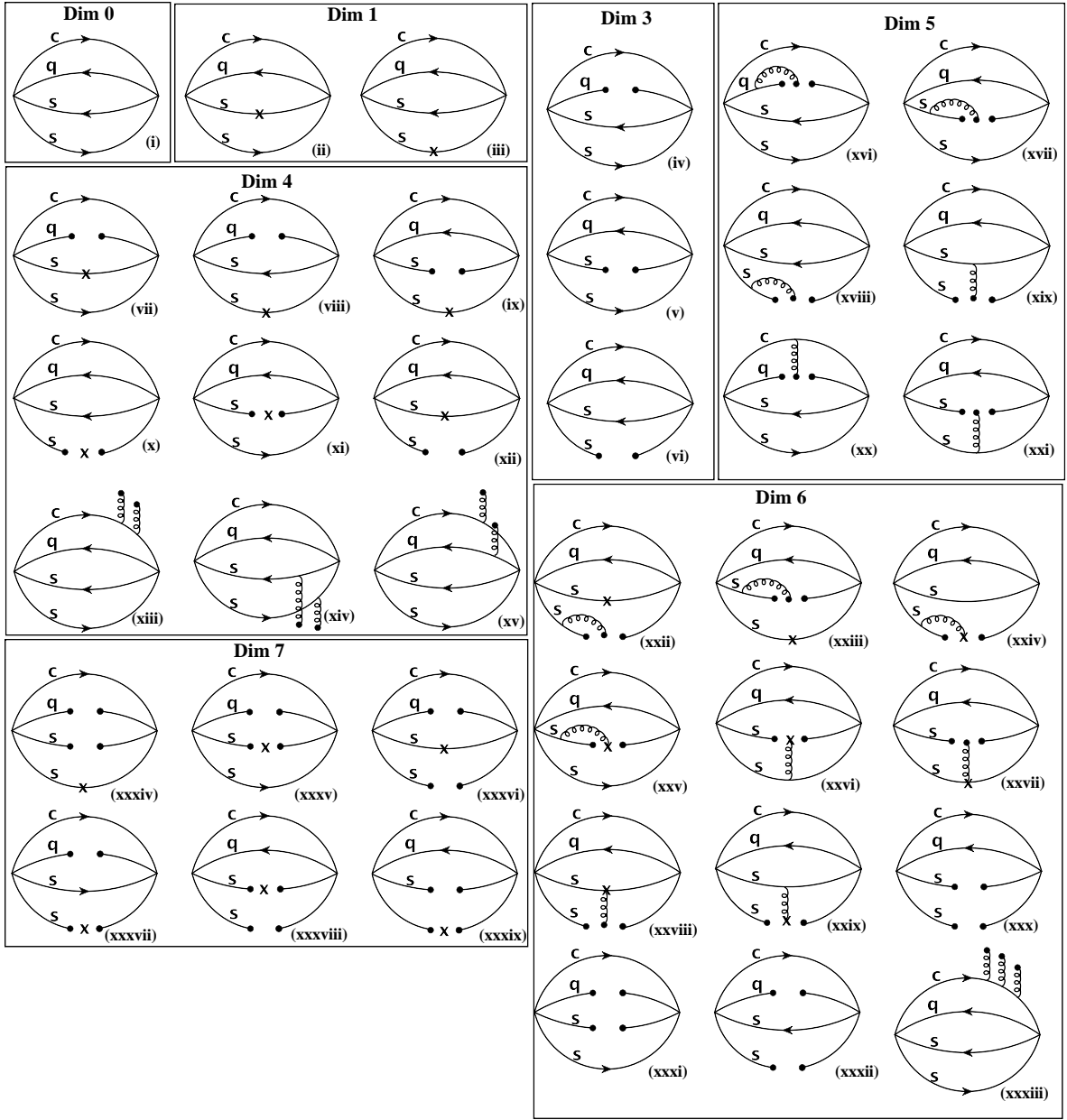}
\caption{Diagrams which contribute to the OPE side of the sum rule.}
\label{graph}
\end{figure}

In the expansion of Eq.~(\ref{ope}), the operators are ordered according to their 
dimension n, where $n = 0$ corresponds to the unit operator, i.e., perturbative 
contribution, and the rest of operators are related to the QCD vacuum fields in
 terms of condensates. For  normal quark-antiquark states, the contributions of 
condensates with dimension higher than four are suppressed by large powers of 
$\Lambda^2_{QCD}/Q^2$, with $1/\Lambda_{QCD}$ being the typical long-distance scale.
 However, for molecular states, condensates with higher dimensions can play an 
important role. This is taken into account by writing Eq.~(\ref{ope}) in terms of 
the spectral density using the dispersion relation

\begin{equation}
\Pi^{\rm OPE} \left( q^2 \right) = \int\limits_{m_c^2}^{\infty} ds\,\, \frac{\rho^{\rm OPE}(s)}{s - q^2} \,+ \textrm {  Subtraction terms}.\label{pi_1}
\end{equation}
 We work at leading order in $\alpha_s$ and we consider condensates up to dimension
seven, as shown in Fig.~\ref{graph}. 

Therefore, $\rho^{\rm OPE}$ can be written as:
\begin{equation}
\rho^{\rm OPE}(q^2)  = \rho^{pert} + \rho^{m_s}  + \rho^{\langle \bar{q} q \rangle} + \rho^{\langle g^2 G^2\rangle} +  \rho^{m_s \langle \bar{q} q \rangle} + \rho^{\langle \bar{ q}  g \sigma \cdot G q \rangle} + \rho^{m_s \langle \bar{ q}  g \sigma \cdot G q \rangle} + \rho^{\langle \bar{q} q \rangle^2} + \rho^{m_s \langle \bar{q} q \rangle^2},\label{rho}
\end{equation}
where $m_{s}$ represents the mass of the strange quark. The spectral density 
$\rho^{\rm OPE}$ is  related to the imaginary part of the correlation function as 
$\pi \rho^{\rm OPE} (s) =$ Im $\bigl[ \Pi^{\rm OPE} (s) \bigr]$.

To calculate the different terms in Eq.~(\ref{rho}),  for the $D_{s^*0} \bar{K}$ 
and  $D f_0$ currents, we use the momentum-space expression for the heavy quark
propagator and the coordinate-space expression for the light quark propagator.
The Schwinger parameters are used to evaluate the heavy quark part of the correlator
and to perform the $d^4x$ integration in Eq.~(\ref{corfn}). 
Finally we get integrals in the Schwinger parameters. The result of these integrals 
are given in terms of logarithmic functions, from where we extract the spectral densities
and the limits of the integration.

Carrying out the calculations  for the different diagrams shown in Fig.~\ref{graph} leads us to the following expressions, where $m_c$ is the mass of 
the charm quark.
\begin{enumerate}
{\item The perturbative or dimension 0 contribution is found to be: \begin{equation}
\rho^{\rm pert}_{\raisebox{-4pt}{\scriptsize $D_{s^*0} \bar{K}$}}  (q^2) = \rho^{pert}_{\raisebox{-4pt}{\scriptsize $D f_0$}}  (q^2)= -\int\limits_{0}^{\alpha_{max}} d\alpha\, \frac{\left((\alpha -1) q^2+m^2_c\right)^4 \alpha ^3}{2^{12} \pi ^6 (\alpha -1)^3}.\label{ope0}
\end{equation}}

\item
For the terms of dimension 1, which are proportional to $m_s$, we get:
\begin{eqnarray}\nonumber
\rho^{m_s}_{\raisebox{-4pt}{\scriptsize $D_{s^*0} \bar{K}$}}  (q^2) &=& -\int\limits_{0}^{\alpha_{max}} d\alpha\, \frac{m_c \left((\alpha -1) q^2+m^2_c\right)^3 \alpha ^3 m_s}{2^{10} \pi ^6 (\alpha -1)^3},\\
\rho^{m_s}_{\raisebox{-4pt}{\scriptsize $D f_0$}}  (q^2) &=& 0.\label{ope1}
\end{eqnarray}

\item
The calculation of the diagrams with one quark condensate gives
\begin{eqnarray}\nonumber
\rho^{\langle \bar{q}\,q \rangle}_{\raisebox{-4pt}{\scriptsize $D_{s^*0} \bar{K}$}} (q^2)&=&\int\limits_{0}^{\alpha_{max}} d\alpha\,\frac{3 m_c \left((\alpha -1) q^2+m^2_c\right)^2 \alpha ^2 \langle \bar{s}\,s\rangle }{2^{8} \pi ^4 (\alpha -1)^2},\\
\rho^{\langle \bar{q}\,q \rangle}_{\raisebox{-4pt}{\scriptsize $D f_0$}} (q^2)&=& -\int\limits_{0}^{\alpha_{max}} d\alpha\, \frac{3 m_c \left((\alpha -1) q^2+m^2_c\right)^2 \alpha ^2 \langle \bar{q}\,q\rangle }{2^{8} \pi ^4 (\alpha -1)^2}.\label{ope2}
\end{eqnarray}

\item
Both $\rho^{m_s \langle \bar{q}\,q \rangle}$ and $\rho^{\langle g^2 G^2 \rangle}$ 
contribute to dimension four and the expressions for the corresponding spectral 
densities are,
\begin{eqnarray}\nonumber
\rho^{m_s \langle \bar{q}\,q \rangle}_{\raisebox{-4pt}{\scriptsize $D_{s^*0} \bar{K}$}} (q^2)&=&\int\limits_{0}^{\alpha_{max}} d\alpha\, \frac{3 \left((\alpha -1) q^2+m^2_c\right)^2 \alpha   m_s}{2^{7} \pi ^4 (\alpha -1)} \biggl[  \langle \bar{q}\,q\rangle -  \langle \bar{s}\,s\rangle \biggr],\\
\rho^{m_s \langle \bar{q}\,q \rangle}_{\raisebox{-4pt}{\scriptsize $D f_0$}} (q^2)&=& -\int\limits_{0}^{\alpha_{max}} d\alpha\, \frac{9 \left((\alpha -1) q^2+m^2_c\right)^2 \alpha  \langle \bar{s}\,s\rangle  m_s}{2^{7} \pi ^4 (\alpha -1)},\label{ope3}
\end{eqnarray}
\begin{eqnarray}\nonumber
\rho^{\langle g^2 G^2 \rangle}_{\raisebox{-4pt}{\scriptsize $D_{s^*0} \bar{K}$}} (q^2) = 
\rho^{\langle g^2 G^2 \rangle}_{\raisebox{-4pt}{\scriptsize $D f_0$}} (q^2) &=& \int\limits_{0}^{\alpha_{max}} d\alpha\, \frac{3\langle g^2 G^2\rangle}{2^{12}\pi^6}\left(\frac{(2-\alpha)(m^2_c+(\alpha-1)q^2)}{2(1-\alpha)}+\frac{m^2_c\alpha^2}{9(1-\alpha)^2}\right)\\
&&\times \frac{\alpha(m^2_c+(\alpha-1)q^2)}{1-\alpha}\label{ope4}
\end{eqnarray}

\item Considering the mixed condensates, we get
\begin{eqnarray}\nonumber
\rho^{\langle \bar{q} g\sigma\cdot G q \rangle}_{\raisebox{-4pt}{\scriptsize $D_{s^*0} \bar{K}$}} (q^2)&=& \int\limits_{0}^{\alpha_{max}} d\alpha\, \frac{3 m_c\langle \bar s g\sigma\cdot Gs\rangle\alpha(1-2\alpha)(m^2_c+(\alpha-1)q^2)}{2^{8}\pi^4(\alpha-1)^2},\\
\rho^{\langle \bar{q} g\sigma\cdot G q \rangle}_{\raisebox{-4pt}{\scriptsize $D f_0$}} (q^2)&=&-\int\limits_{0}^{\alpha_{max}} d\alpha\, \frac{3 m_c\langle \bar q g\sigma\cdot Gq\rangle\alpha(1-2\alpha)(m^2_c+(\alpha-1)q^2)}{2^{8}\pi^4(\alpha-1)^2}\label{ope5}
\end{eqnarray}

\item Going to the dimension 6 operator, we get the following contributions for 
the terms proportional to $m_s \langle \bar{q} g\sigma\cdot G q \rangle$,
\begin{eqnarray}\nonumber
\rho^{m_s \langle \bar{q} g\sigma\cdot G q \rangle}_{\raisebox{-4pt}{\scriptsize $D_{s^*0} \bar{K}$}} (q^2)&=&\int\limits_{0}^{\alpha_{max}} d\alpha\, \frac{m_s }{2^8 \pi^4} \left[ m_c^2 - q^2 \left(1- \alpha \right) \right]\\\nonumber
 &&\left( \langle \bar{q} g \sigma \cdot G q \rangle \left(6 ln \left( \alpha \right)  - 3 \right) - \langle \bar{s} g \sigma \cdot G s \rangle \left(\frac{1 + 2 \alpha}{1 - \alpha}\right) \right),\\
\rho^{m_s \langle \bar{q} g\sigma\cdot G q \rangle}_{\raisebox{-4pt}{\scriptsize $D f_0$}} (q^2)&=& \int\limits_{0}^{\alpha_{max}} d\alpha\, \frac{m_s  \langle \bar{s} g \sigma \cdot G s\rangle}{2^7 \pi^4} \left[ m_c^2 - q^2 \left(1- \alpha \right) \right] \left(1 -  6 ln \left( \alpha \right) \right),\label{ope6}
\end{eqnarray}
 four-quark condensates
\begin{eqnarray}\nonumber
\rho^{\langle \bar{q}\,q \rangle^2}_{\raisebox{-4pt}{\scriptsize $D_{s^*0} \bar{K}$}} (q^2)=- \int\limits_{0}^{\alpha_{max}} d\alpha\, \frac{\left((\alpha -1) q^2+m^2_c\right) \langle \bar{q}\,q\rangle \langle \bar{s}\,s\rangle}{2^{4} \pi ^2},\\
\rho^{\langle \bar{q}\,q \rangle^2}_{\raisebox{-4pt}{\scriptsize $D f_0$}} (q^2)= \int\limits_{0}^{\alpha_{max}} d\alpha\, \frac{\left((\alpha -1) q^2+m^2_c\right) \langle \bar{s}\,s\rangle ^2}{2^{4} \pi ^2},\label{ope7}
\end{eqnarray}
and  three-gluon condensates
\begin{equation}
\rho^{\langle g^3 G^3 \rangle}_{\raisebox{-4pt}{\scriptsize $D_{s^*0} \bar{K}$}} (q^2)= \rho^{\langle g^3 G^3 \rangle}_{\raisebox{-4pt}{\scriptsize $D f_0$}} (q^2)= - \int\limits_{0}^{\alpha_{max}} d\alpha\, \frac{\left((\alpha -1) q^2+3 m^2_c\right) \alpha ^3 \langle g^3 G^3\rangle }{3\times2^{14} \pi ^6 (\alpha -1)^3}. \label{ope8}
\end{equation}
In the case of the (dimension six) four-quark condensate, we have used the
factorization  assumption. Therefore, its  vacuum saturation value 
is given by:
\begin{equation}
\langle\bar{q}q\bar{q}q\rangle=\langle\bar{q}q\rangle^2.
\end{equation}

\item Finally, for dimension 7, we get
\begin{eqnarray}\nonumber
\rho^{m_s \langle\bar{q}\,q \rangle^2}_{\raisebox{-4pt}{\scriptsize $D_{s^*0} \bar{K}$}} (q^2)&=& \int\limits_{0}^{\alpha_{max}} d\alpha\, \frac{m_c  m_s}{2^{3} \pi ^2} \left( \frac{\langle \bar{s}\,s\rangle^2}{2^2} - \langle \bar{q}\,q\rangle  \langle \bar{s}\,s\rangle \right),\\
\rho^{m_s \langle\bar{q}\,q \rangle^2}_{\raisebox{-4pt}{\scriptsize $D f_0$}} (q^2)&=& - \int\limits_{0}^{\alpha_{max}} d\alpha\, \frac{3 m_c \langle \bar{q}\,q\rangle  \langle \bar{s}\,s\rangle  m_s}{2^{4} \pi ^2}\label{ope9}
\end{eqnarray}

\end{enumerate}
The integration limit in Eqs.(\ref{ope0})-(\ref{ope9}) is $\alpha_{max} = 1 - \frac{m_c^2}{q^2}$. For numerical calculations we need the values of the different 
condensates and quark masses. We have used here the same values for these inputs as those used in QCDSR 
calculations for other exotic molecular states \cite{mfrev,matheus,bracco,narison}, 
which are given in Table~\ref{parameters}. For the $\langle g^3 G^3 \rangle$ condensate,
we have used the new numerical value estimated in Ref.~\cite{narison2}.

\begin{table}[h!]
\caption{Values of the different known parameters required for numerical 
calculations of the correlation function given by Eq.~(\ref{corfn}) 
(see Refs.~\cite{mfrev,matheus,bracco,narison,narison2}).}\label{parameters}
\begin{ruledtabular}
\begin{tabular}{cc}
Parameters & Values\\
\hline
$m_s$& $0.10 \pm 0.022$ GeV\\
$m_c$& $1.23 \pm 0.05$ GeV\\
$\langle \bar{q} q \rangle$ & $-(0.23 \pm 0.03)^3$ GeV$^3$\\
$\langle \bar{s} s \rangle$ & 0.8 $\langle \bar{q} q \rangle$\\
$\langle g^2 G^2 \rangle$ & $(0.88\pm0.25)$ GeV$^4$\\
$\langle g^3 G^3 \rangle$ & $(0.58\pm0.18)$ GeV$^6$\\
$\langle \bar{q} \sigma \cdot G q \rangle$& 0.8$\langle \bar{q} q \rangle$ GeV$^2$\\
\end{tabular}
\end{ruledtabular}
\end{table}

We now calculate the correlation function from the hadronic or  phenomenological 
point of view.  In this case, the currents  $j^\dagger$ and $j$ are interpreted as 
the creation and annihilation operator of the hadrons which have the quantum numbers
 of the current $j$. For this $\Pi (q^2)$ is written  by inserting a complete set
of states  with the same quantum numbers as  those of the currents under 
consideration 
\begin{equation}
\Pi^{\rm phenom} (q^2) = i \int d^4x e^{i q\cdot x} \int\frac{d^3p}{2 p^0 \left( 2\pi \right)^3}\sum\limits_{k=0}^\infty \langle 0 \mid j (x) \mid m_k \vec{p}\rangle\langle m_k \vec{p}\mid j^{\dagger}(0) \mid 0 \rangle. \label{corfn2}
\end{equation}
Thus, the correlation function contains the information on all the hadrons of a 
given set of quantum numbers including the one we are interested in, which is the 
low mass, relatively narrow, hadron of the series.  One proceeds in such a situation
 by assuming that the spectral density of hadrons, for a fixed set of quantum 
numbers, can be expressed as a sum of a narrow, sharp state (which we are interested
 in), and a smooth continuum
\begin{equation}
\rho^{\rm phenom}(s) = \lambda^2 \delta(s - m^2) +  \rho_{\rm continuum} (s), \label{rho_phenom}
\end{equation}
where $\rho_{\rm continuum}$ is assumed to vanish below a certain value of $s$, 
$s_{0}$, which corresponds to the continuum threshold. Above this threshold, it is 
assumed to be given by the result obtained with the OPE. Therefore, one uses the 
ansatz \cite{io1} $\rho_{\rm continuum} (s)=\rho^{\rm OPE}(s) \Theta(s-s_{0})$.

The delta function in Eq.~(\ref{rho_phenom}) implies that the width of the particle is 
assumed to be zero. In principle, the introduction  of a finite  
width in the above calculation could change the final result obtained for the mass and,
 more importantly, it could be another important source of error in the final result 
for  the mass. However, our experience with this type of calculation suggests that the 
introduction of a width is not a very important source of errors.  Indeed, in  
Ref.~\cite{width1} (see also the discussion in  Ref.~\cite{mfrev}) a careful discussion of this 
effect was presented with the conclusion that for the X(3872), Z(4430) and Z(4250) the 
uncertainty in the width, when properly taken into account generates  at most a 5\% 
error in the final mass of the state. Moreover, in Ref.~\cite{widht2} a careful study of the 
role played by the particle width was performed. The semileptonic decay 
$D \rightarrow \kappa  l \nu$ was calculated with QCDSR. From experiment we
know that $m_{\kappa} = 0.797$ GeV and the width is $\Gamma_{\kappa} = 0.410$ GeV. 
With this extremely large value of the width, we would expect that the zero width 
approximation for the $\kappa$ would change the result dramatically. However, as shown 
in the quoted article, the zero width approximation yields a total $D$ semileptonic 
decay rate which is only about 20 \% larger. Given the huge size of the kappa width 
(half of its mass!), the above mentioned estimate could be considered an upper limit 
of the error introduced by neglecting the particle width. In view of these examples 
and bearing in mind the exploratory nature of the present work, we will postpone the 
inclusion of the width for a future study.
However, we are aware and must remind the reader that the 
estimated error in our results could be slightly larger.

In Eq.~(\ref{rho_phenom}) $\lambda$ is the coupling of the current $j$ with the 
low-lying hadron with mass $m$: $ \langle 0 | j| m \rangle =  \lambda$.

The spectral density given by Eq.~(\ref{rho_phenom}) is related to the correlation 
function of Eq.~(\ref{corfn2}) as
\begin{equation}
\Pi^{\rm phenom} (q^2) = \frac{\lambda^2}{m^2 - q^2}+ \int\limits_{s_0}^{\infty} ds \frac{\rho^{\rm OPE}(s)}{s - q^2}, \label{pi_2}
\end{equation}

To carry out the calculations, $s_0$ is taken as a parameter of the method but its 
value is not completely arbitrary: it is related  to the onset of the continuum in 
the current $j$ under consideration and is taken to be roughly 0.5 GeV above the 
mass of the  hadron we are interested in~\cite{colangelo2,mfrev}. In this work, we 
are looking for a resonance with a possible $D f_0(980)$ or $D_{s^*0}(2317) \bar{K}$ molecule-like 
structure. Since such resonances are weakly bound, they are expected to get 
generated close to the threshold of the constituent mesons. Thus, $\sqrt{s}_0$ in 
the present case can be $\sim$ 3.4 GeV.

The correlation function calculated using QCD suffers from divergent contributions 
coming from long range interactions, while the one calculated phenomenologically 
contains contribution from the continuum.  This situation can be  improved by taking
 Borel transform of both Eqs.~(\ref{pi_1}) and (\ref{pi_2}), which kills the 
problematic terms of both sides, and which is defined as:
\begin{equation}
{\cal B}_{M^2}[\Pi(q^2)]=
\lim_{\stackrel{\scriptstyle -q^2,n\rightarrow\infty}
{\scriptstyle -q^2/n=M^2}}
{(-q^2)^{n+1}\over n!}\left(d\over d q^2\right)^n
\Pi(q^2)\;.
\label{borel}
\end{equation}
 
After taking the Borel transform, we equate the resulting expressions of the 
correlation functions on the basis of its dual nature and get
\begin{align}
\lambda^2 e^{-m^2/M^2} &+ \int\limits_{s_0}^{\infty} ds \rho^{\rm OPE} (s) e^{-s/M^2} = \int\limits_{m_c^2}^{\infty} ds\rho^{\rm OPE}(s) e^{-s/M^2},
\end{align}
which can be rearranged as
\begin{align}
\lambda^2 e^{-m^2/M^2}&= \int\limits_{m_c^2}^{s_{0}} ds\rho^{\rm OPE}(s) e^{-s/M^2}, 
\label{equate}
\end{align}
where $M$ represent the Borel mass parameter. Calculating the derivative of 
Eq.~(\ref{equate}) with respect to $M^{2}$ and dividing the resulting expression by 
Eq.~(\ref{equate}), we obtain the mass sum rule
\begin{equation}
m^2 = \dfrac{\int\limits_{m_c^2}^{s_0} ds \, s \, \rho^{\rm OPE} (s) e^{-s/M^2}}{\int\limits_{m_c^2}^{s_0} ds \rho^{\rm OPE} (s) e^{-s/M^2}}.\label{masssr}
\end{equation}
Having the mass one can evaluate the current-state coupling constant through 
Eq.~(\ref{equate})
\begin{equation}
\lambda^2 = \dfrac{\int\limits_{m_c^2}^{s_0} ds \, s \, \rho^{\rm OPE} (s) e^{-s/M^2}}{e^{-m^2/M^2}}. \label{decon}
\end{equation}

The reliability of the results obtained within QCD sum rules depends on the 
definition of a valid Borel window. This range of the Borel mass is obtained by 
making the following constraints: 
\begin{itemize} 
\item The maximum value of the Borel mass, $M_{\rm max}$, where the results should 
be reliable, is fixed by ensuring that the pole term (low mass hadron) gives the 
dominant contribution to the calculations. However, $M_{\rm max}$ is a function of 
$s_0$. As mentioned earlier, a reasonable value of $\sqrt{s}_0$ in the present calculation 
can be 3.4 GeV. We show the  contributions of the pole and continuum terms weighted 
by their sum  \cite{mfrev,matheus,bracco} for  the $D f_0 (980)$  and $D_{s^*0} (2317) \bar{K}$  systems,
 obtained with $\sqrt{s}_0 = 3.4$ GeV, in Fig.~\ref{poleandcont}, which shows that 
$M_{\rm Max}^2 \sim$ 2.06 GeV$^2$  in the former case and  1.79 GeV$^2$ in the latter one, respectively. 
\begin{figure}[ht!]
\includegraphics[width= 0.49\textwidth,height=6cm]{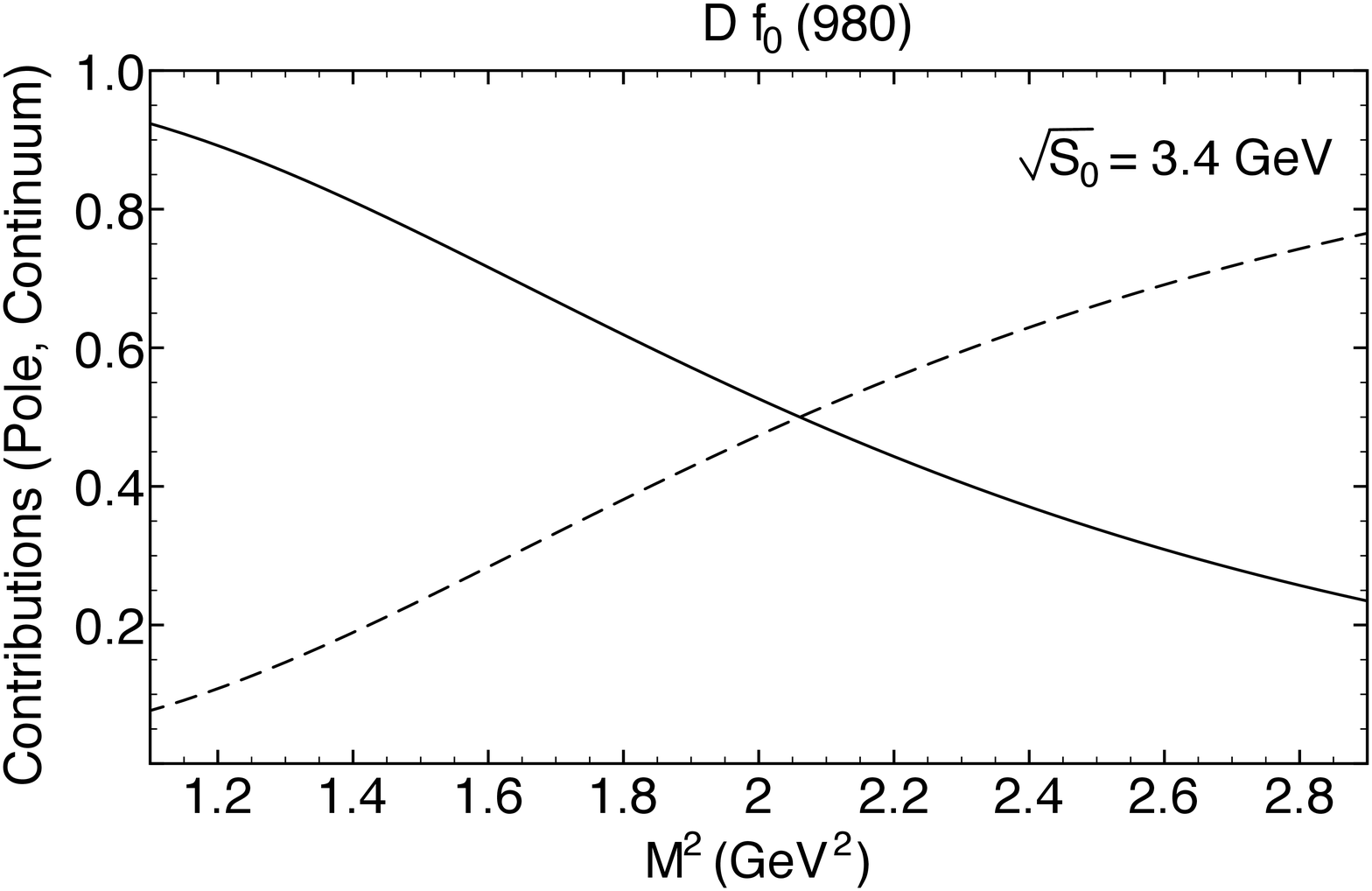}
\includegraphics[width= 0.50\textwidth,height=6cm]{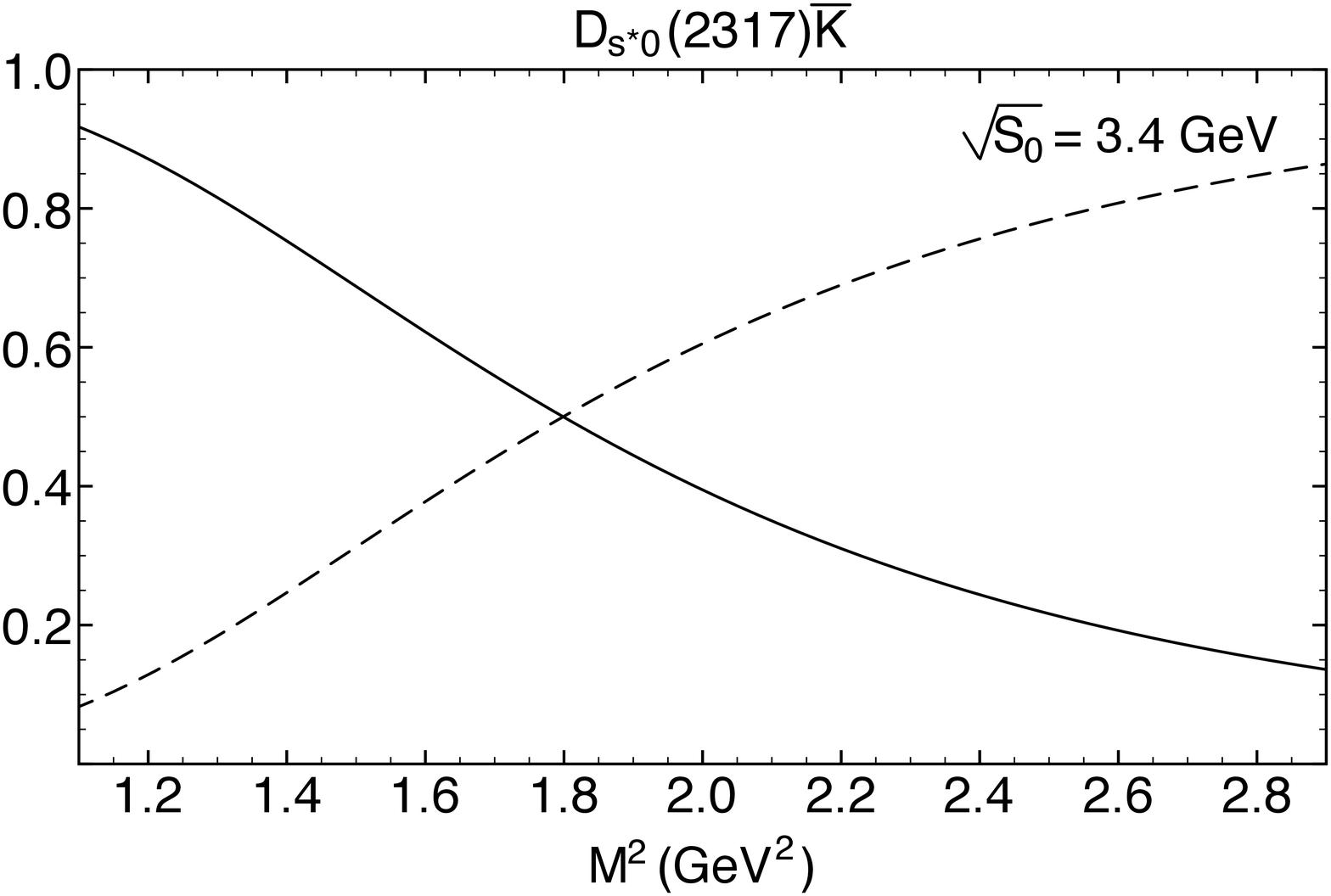}
\caption{The  contributions of the pole (solid line) and continuum (dashed line) weighted by (divided by) their sum for the $D f_0(980)$ (left panel) and $D_{s^*0} (2317) \bar{K}$ 
(right panel) systems. }\label{poleandcont}
\end{figure}

\item The second constraint is to look  for that Borel mass range where  a 
convergence in the OPE series is found. For this we calculate the perturbative 
contribution and add to it the diagrams with higher dimensions step by step.  In 
other words, we calculate the right hand side of Eq.~(\ref{equate}) by first using 
Eq.~(\ref{ope0}) for $\rho^{\rm OPE}$, then by using the sum of Eqs.~(\ref{ope0}) 
and (\ref{ope1}), which means including the diagrams up to dimension 1, next we do 
the calculations up to the subsequent higher dimension by taking a sum of 
Eqs.~(\ref{ope0})-(\ref{ope2}), etc, until going to diagrams with dimension 7 (given
 by Eqs.~(\ref{ope9})). For a convenient comparison, the result obtained in each 
case is weighted (divided) by the one obtained by using the whole series of 
Eq.~(\ref{rho}) for the spectral density. 
 In Fig.~\ref{fig_allterms}, we show the results of such an analysis of OPE 
convergence for the $D f_0(980)$ (left panel) as well as $D_{s^*0} (2317) \bar{K}$ 
(right panel) systems. 
\begin{figure}[htbp!]
\includegraphics[width= 0.49\textwidth,height=6cm]{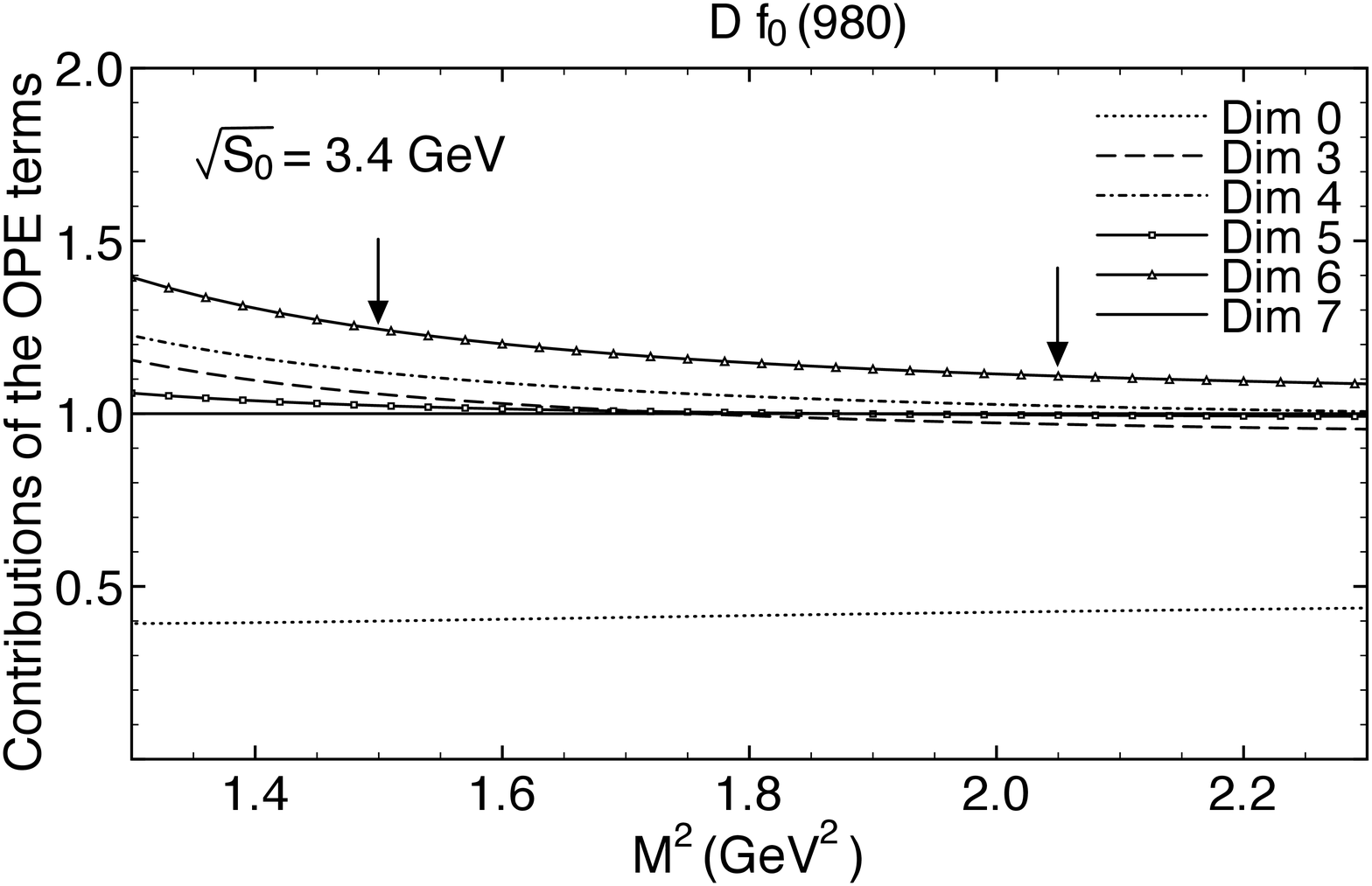}
\includegraphics[width= 0.50\textwidth,height=6cm]{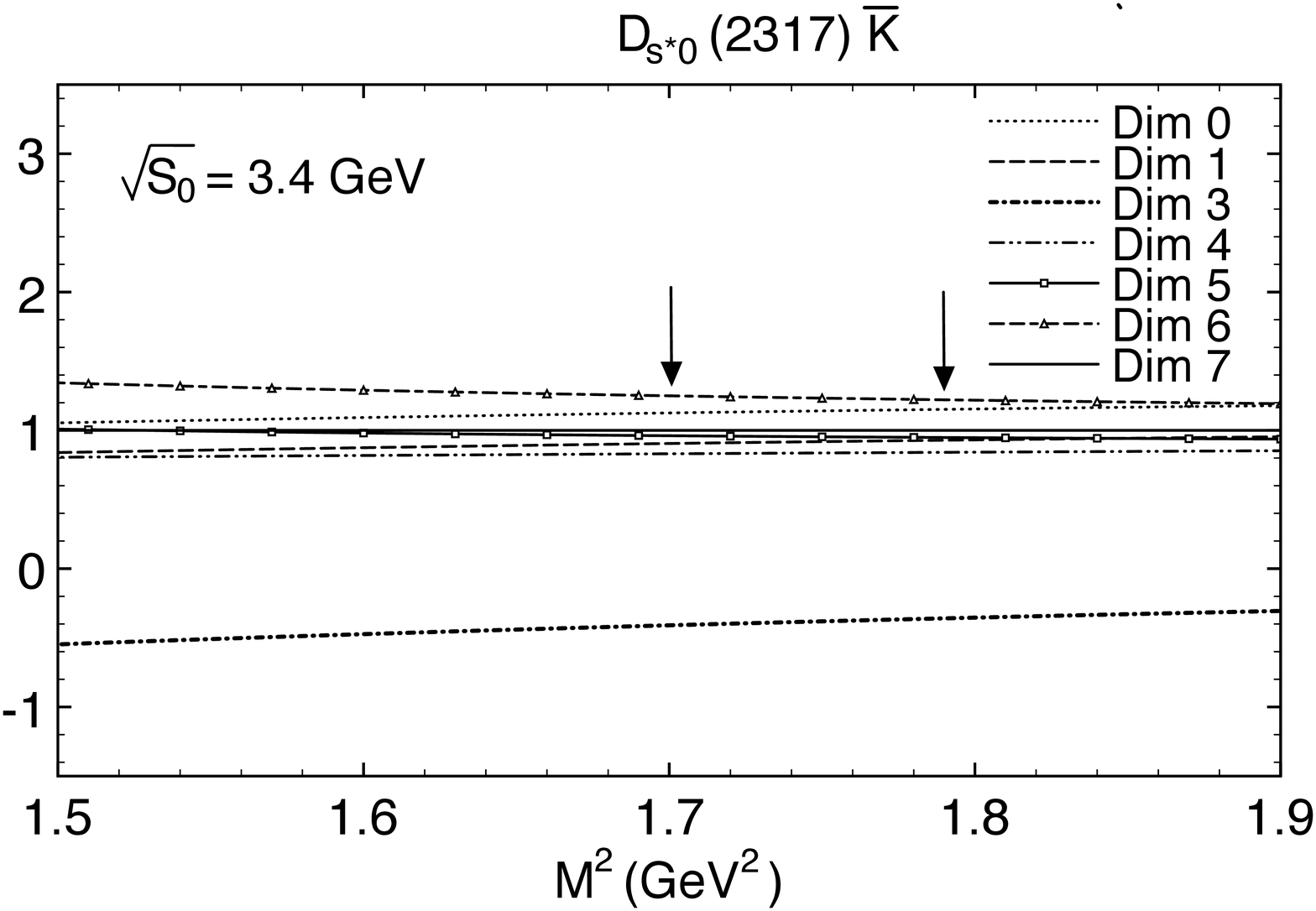}
\caption{Relative contributions of the different OPE terms as a function of the squared Borel Mass, for the $D f_0(980)$ (left panel) as well as $D_{s^*0} (2317) \bar{K}$ (right panel) systems. A value of $\sqrt{s}_0 =$ 3.4 GeV is used in these calculations. The arrows in the figures indicate the valid Borel window, which is determined by using the conditions discussed in the text.}\label{fig_allterms}
\end{figure}

The final condition, which is imposed to identify the minimum value for the Borel 
mass, is that the contribution defined by
\begin{equation}
\left | \frac{\sum\limits_{{\rm dim} = 1}^{\rm N_{max}-1} \left( \int\limits_{m_c^2}^{s_0} ds  \, \rho^{\rm OPE}_{\rm dim} (s) e^{-s/M^2}\right)-\sum\limits_{{\rm dim} = 1}^{\rm N_{max}} \left(\int\limits_{m_c^2}^{s_0} ds  \, \rho^{\rm OPE}_{\rm dim}  (s) e^{-s/M^2}\right)}{\sum\limits_{{\rm dim} = 1}^{\rm N_{max}} \left(\int\limits_{m_c^2}^{s_0} ds  \, \rho^{\rm OPE}_{\rm dim}  (s) e^{-s/M^2}\right)}\right |
\end{equation}
is less than 0.25. In the equation written above, $N_{max}$ refers to the maximum 
dimension of the condensates taken into account in the calculation, which is 7 in 
the present case. 
With this, the OPE convergence is ensured in the Borel mass range where the 
results can be taken as the reliable ones.  It can be see from 
Fig.~\ref{fig_allterms} that a good convergence of the OPE series is found for
$M^2_{Min}=$ 1.5 GeV$^2$ and 1.7 GeV$^2$, for the $D f_0(980)$ (left panel) and 
$D_{s^*0} (2317) \bar{K}$ (right panel) cases respectively.

The valid Borel windows established using both criteria discussed above are indicated
with arrows for  the $D f_0(980)$ and  $D_{s^*0} (2317) \bar{K}$  systems in 
Fig.~\ref{fig_allterms}, for $\sqrt{s}_0 =$ 3.4 GeV.

\end{itemize}

\begin{figure}[ht!]
\includegraphics[width= 0.49\textwidth,height=6cm]{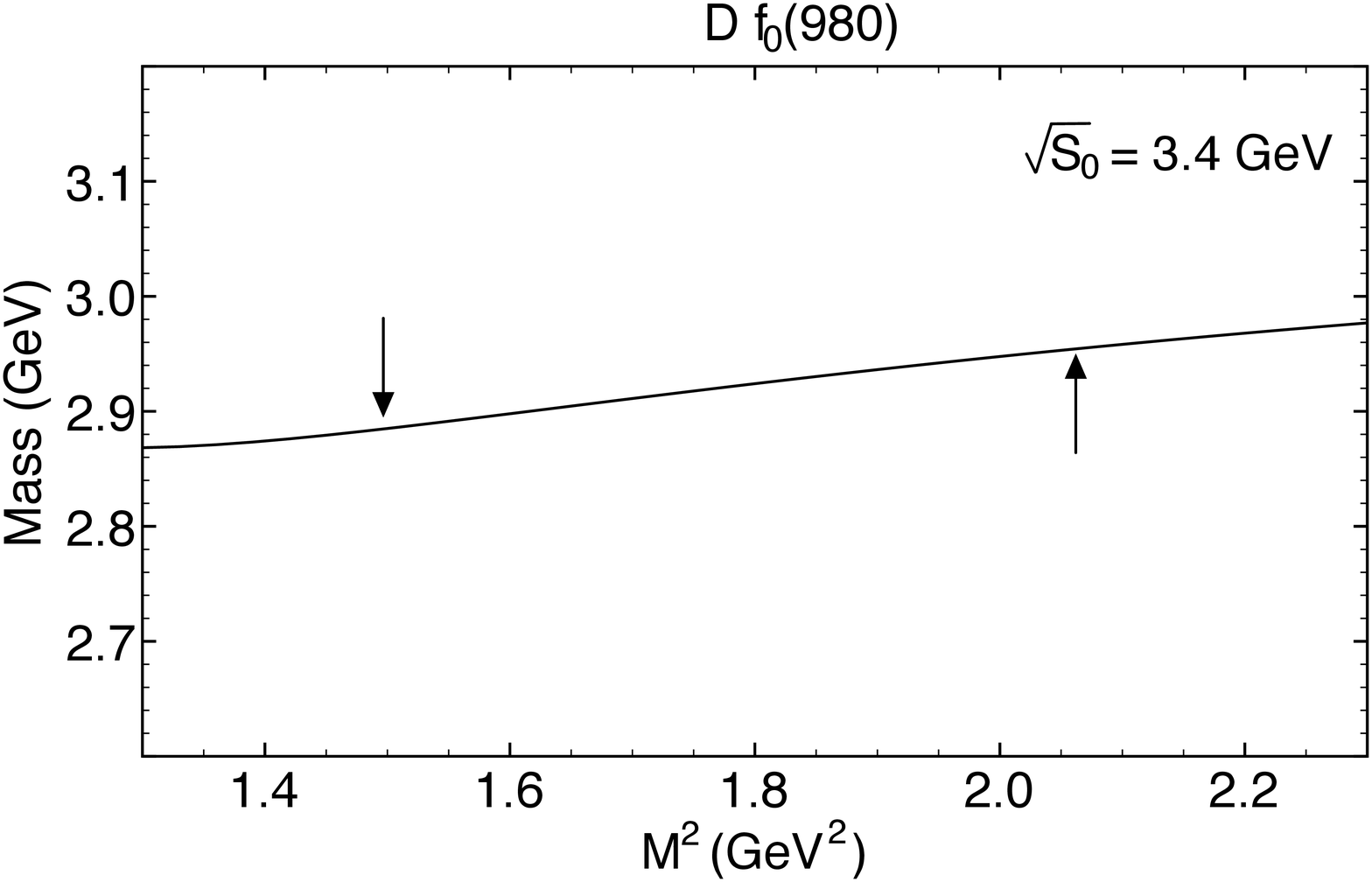}
\includegraphics[width= 0.50\textwidth,height=6cm]{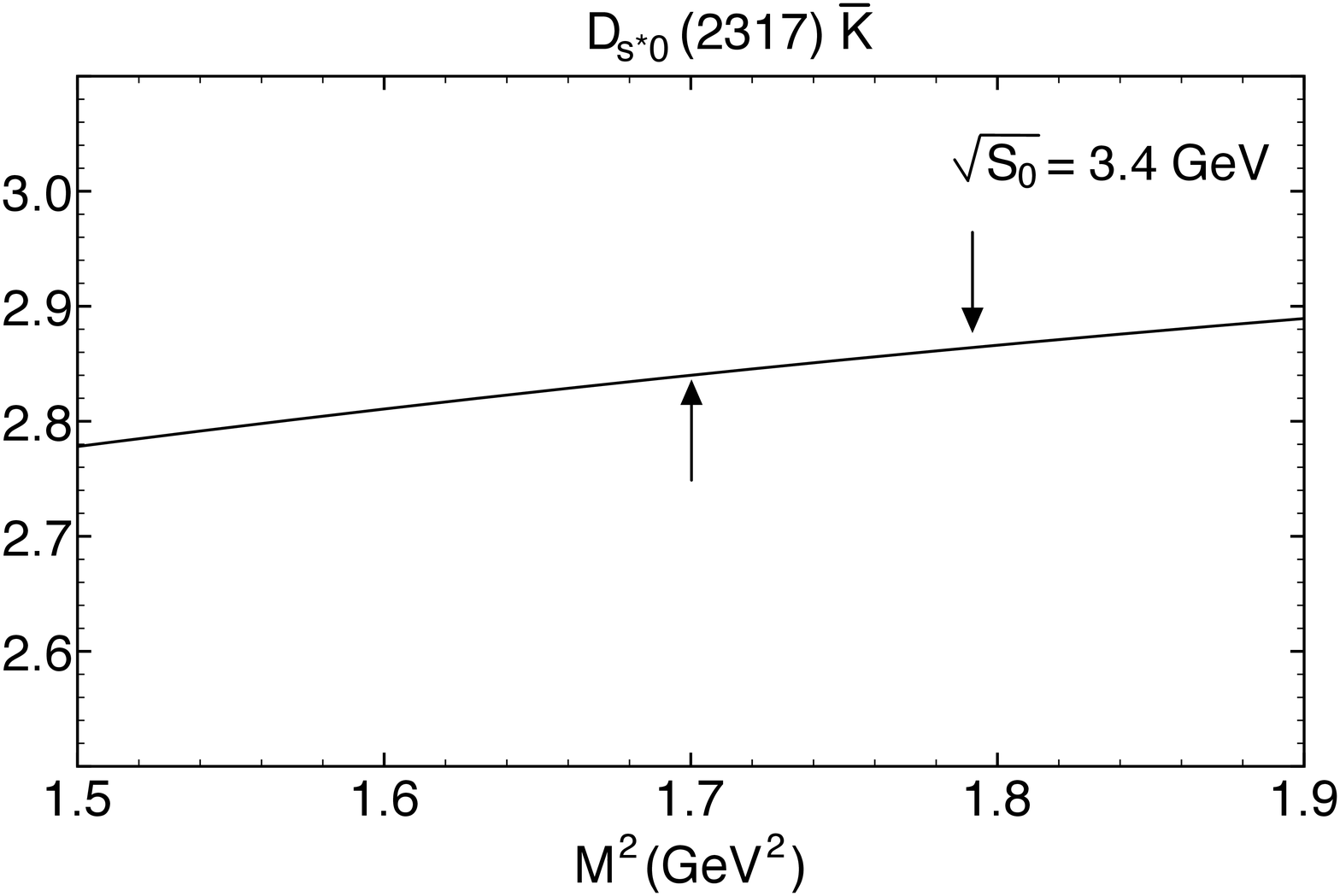}
\caption{Mass of a molecule-like resonance obtained by using QCD sum-rules to study the $D f_0(980)$ (left panel) and $D_{s^*0} (2317) \bar{K}$ (right panel) systems. The arrows indicate the range where the reliability of the results is ensured. }\label{mass_fig}
\end{figure}

Having fixed these conditions, we show, in Fig.~\ref{mass_fig}, the results of the 
calculation of the mass (using Eq.~(\ref{masssr})) of the states described using 
$D f_0(980)$ and $D_{s^*0} (2317) \bar{K}$ molecule-like currents. 
It can be seen that a mass of 2.852 $\pm$  0.008  GeV is found in the case of 
$D_{s^*0} (2317) \bar{K}$ while it is  2.921 $\pm$  0.021 GeV in the $D  f_0(980)$ system 
(within the Borel window, indicated by the arrows in Fig.~\ref{mass_fig}).  
However, these calculations have been done using  the value of  3.4 GeV for 
$\sqrt{s}_0$, which is a parameter. We now vary the continuum threshold in the 
range $3.3\leq\sqrt{s_0}\leq3.6$ GeV to check the sensitivity of our results to 
this parameter.
 We also take into account the fact that there exists  uncertainty in the 
knowledge of the values of the different condensates and quark masses listed in 
Table.~\ref{parameters}. Considering all these uncertainties we finally get
\begin{equation}
m_{D_{s^*0}\bar{K}}= ( 2.913  \pm  0.140 ) \mbox{ GeV},
\end{equation} 
for the $D_{s^*0}(2317) \bar{K}$ molecular current, while for the 
$D f_0 (980)$ molecular current we get 
\begin{equation}
m_{Df_0}=( 2.926 \pm  0.237 )\mbox{ GeV}.
\end{equation}
The above results have been determined by averaging the mass over the corresponding Borel windows  and by calculating the standard
deviation to estimate the error. 

Next,  following the same procedure, we have calculated the coupling $\lambda$ for the two configurations studied 
and found that the current-state coupling of the state described by the 
$D_{s^*0} (2317) \bar{K}$ current is around two times weaker than the one found for the
 $D f_0(980)$ current:
\begin{eqnarray}\nonumber
\lambda_{D_{s^*0}\bar{K}} &=&  ( 5.8 \pm 1.2 ) \times 10^{-3}  \mbox{ GeV}^5, \mbox{ and }\\
\lambda_{Df_0}&=& ( 9.4 \pm  3.3) \times 10^{-3}   \mbox{ GeV}^5.\label{coups}
\end{eqnarray}
At a first sight these couplings may look compatible within error bars. However, this is not the case. Our error analysis 
shows that, 
for a given set of parameters, $\lambda_{Df_0}$ turns out to be 1.4~$\lambda_{D_{s^*0}\bar{K}}$~$-$~2~$\lambda_{D_{s^*0}\bar{K}}$ and this situation repeats for all other set of inputs.
We can interpret this result as an indication that a $D f_0(980)$ state is better 
represented by the respective molecular current than the $D_{s^*0} (2317) \bar{K}$
state.

\section{Three-hadron approach}

Let us now discuss the study of the $D K \bar K$ system within a very different 
approach which is based on effective field theories, treating hadrons as the degrees
 of freedom instead of quarks, and examine if  the findings obtained in such a 
calculation are compatible with the ones found with QCD sum rules.

In the unitary chiral models
\cite{kaiser,osetramos,ollerulf,jido3,jido2,oller,pelaez,ollerprog,hyodo,bruns}, the $f_{0}(980)$ 
resonance is described as a molecular hadron state  generated in the interaction 
of the $K\bar K$, and $\pi\pi$ coupled channels~\cite{oller,pelaez}. Similarly, the 
$D_{s^{*} 0} (2317)$ state can be interpreted as a $DK$ bound state formed in the 
$DK$, $D_{s}\eta$ coupled channel system
\cite{kolomeitsev,guozou,daniel,guo2,hanhart}.  In these models, Lagrangians based 
on symmetries like chiral~\cite{gasser,meissner,ecker} and heavy quark symmetries
\cite{burdman,jenkins,yan} are used to determine the lowest order amplitude 
describing the transition between the different coupled channels. These amplitudes 
are further unitarized by using them as driving terms in the Bethe-Salpeter equation,
 and  the scattering matrix $t$ for the system is obtained. Recently, these models 
based on effective field theories, chiral symmetry and unitarity in coupled channels
 have been further extended to investigate the interaction of three-hadron systems  
formed by different mesons and baryons, like, $\pi\bar K N$, $N K\bar K$, $J/\psi 
K\bar K$, $\phi K\bar K$, etc.,
and generation of several hadron states like $\Sigma(1660)$, $\phi(2170)$, $Y(4260)$,
 $N^{*}(1710)$, has been found~\cite{mko1,mko2,mko3,mko4}. Analogously to the 
two-body studies where the scattering matrix is obtained by solving the 
Bethe-Salpeter equation taking as kernel the lowest-order chiral amplitude, in the 
case of the approach of Refs.~\cite{mko1,mko2,mko3,mko4}, the Faddeev equations
\cite{faddeev} are solved, having as driving term the chiral two-body scattering 
matrices for the different pairs of the system. In this way, the input two-body $t$
 matrices in the Faddeev equation contain the information related to the generation 
of the corresponding two-body resonances.

In line with the above mentioned works, a different strategy to the one discussed 
in the previous section to study the $Df_0 (980)$ and $D_{s} \bar K$ systems would 
be to consider $f_{0}(980)$ and $D_{s^{*}0}(2317)$ as molecular resonances formed, 
respectively, in the $K\bar K$ and $D K$ systems together with their respective 
coupled channels and study the three-body system $DK\bar K$ following the approach 
of Refs.~\cite{mko1,mko2,mko3,mko4}. To do this, we consider 10 coupled channels for 
total charge zero and charm $C=1$: $D^{0}K^{+}K^{-}$, $D^{0}K^{0}\bar K^{0}$, $D^{0}
\pi^{+}\pi^{-}$, $D^{0}\pi^{-}\pi^{+}$, $D^{0}\pi^{0}\pi^{0}$, $D^{0}\pi^{0}\eta$, 
$D^{+}K^{0}K^{-}$, $D^{+}\pi^{-}\pi^{0}$, $D^{+}\pi^{-}\eta$, $D^{+}\pi^{0}\pi^{-}$.

As mentioned above, to solve the Faddeev equations for the $DK\bar K$ system and 
coupled channels, we first need to determine the two-body scattering matrices $t$ 
for the different pairs of the system. This is done by solving the Bethe-Salpeter 
equation through its on-shell factorization form
\cite{oller,osetramos,pelaez,ollerprog,hyodo}, 
\begin{align}
t=(1-V\mathcal{G})^{-1}V. \label{BS}
\end{align}
The kernel $V$ in Eq.~(\ref{BS})  corresponds to the lowest order two-body amplitude
 obtained from a suitable Lagrangian and $\mathcal{G}$ represents the loop function 
of two hadrons. 

In case of the $DK\bar K$ system and coupled channels, we have two different 
types of interactions: one involving two light pseudoscalars, like $K\bar K$, 
$\pi\pi$, and the other between a heavy and a light pseudoscalar meson, like $DK$, 
$D\pi$.

For the description of the $K\bar K$ system, we follow Refs.~\cite{oller,pelaez} 
and solve Eq.~(\ref{BS}) considering $K\bar K$, $\pi\pi$ and $\pi\eta$ as coupled 
channels. The kernel $V$ is obtained from
the lowest order chiral Lagrangian for the process $PP\to PP$, with $P$ representing
a light pseudoscalar (i.e., $\pi$, $K$, $\eta$)

\begin{align}
\mathcal{L}_{PP}=\frac{1}{12 f^{2}} \textrm {\large Tr} \biggl\{ (\partial_{\mu}PP-P\partial_{\mu}P)^{2}+MP^{4} \biggr\}\label{lchiral}.
\end{align}
In Eq.~(\ref{lchiral}), $f$ is the pion decay constant,  Tr$\{... \}$
indicates the trace in the flavor space of the SU(3) matrices appearing 
in $P$,  which is a matrix containing the different Goldstone bosons and $M$ a mass 
matrix: 
\begin{align}
P=\left(\begin{array}{ccc}\frac{1}{\sqrt{2}}\pi^0+\frac{1}{\sqrt{6}}\eta & \pi^+ & {K}^+ \\ \pi^- & -\frac{1}{\sqrt{2}}\pi^0+\frac{1}{\sqrt{6}}\eta & {K}^0 \\{K}^- & \bar{K}^{0} & -\frac{2}{\sqrt{6}}\eta\end{array}\right),\label{phi}
\end{align}
\begin{align}
M=\left(\begin{array}{ccc}m^2_\pi &0&0\\0& m^2_\pi& 0\\0&0&2m^2_K-m^2_\pi\end{array}\right).\label{mass}
\end{align}
The $V$ matrix obtained using the Lagrangian of Eq.~(\ref{lchiral}) is a function 
of the Mandelstam variables $s$, $t$ and $u$.  This matrix is further projected on 
s-wave and the resulting expressions can be found in Ref.~\cite{oller}.

The loop function $\mathcal{G}$ in Eq.~(\ref{BS}) is calculated using the 
dimensional regularization scheme of Ref.~\cite{pelaez}. In the present case, i.e., 
for a two pseudoscalar system, 

\begin{align}
\mathcal{G}_r=&\frac{1}{16\pi^2}\Bigg\{a_r(\mu)+\ln\frac{m^2_{1r}}{\mu^2}+\frac{m^2_{2r}-m^2_{1r}+E^2}{2E^2}\ln\frac{m^2_{2r}}{m^2_{1r}}\nonumber\\
&+\frac{q_r}{E}\Bigg[\ln\Big(E^2-(m^2_{1r}-m^2_{2r})+2q_rE\Big)+\ln\Big(E^2+(m^2_{1r}-m^2_{2r})+2q_rE\Big)\nonumber\\
&-\ln\Big(-E^2+(m^2_{1r}-m^2_{2r})+2q_rE\Big)-\ln\Big(-E^2-(m^2_{1r}-m^2_{2r})+2q_rE\Big)\Bigg]\Bigg\}.\label{g}
\end{align}
In Eq.~(\ref{g}), $E$ is the total energy of the two-body system, $m_{1r}$, $m_{2r}$
 and $q_r$ correspond, respectively, to the masses and the center of mass momentum 
of the two pseudoscalars present in the r{\it th} channel, $\mu$ is a regularization
 scale and $a_r(\mu)$ a subtraction constant. Following  Ref.~\cite{pelaez}, we have
 taken $\mu =1224$ MeV  and a
value for $a_r(\mu)\sim-1$ (note that there is only one independent parameter here since a change in $\mu$ can be reabsorbed in $a_r$). In this way we can reproduce the observed two-body phase
 shifts and inelasticities for the different  coupled channels as done in 
Refs.~\cite{oller,pelaez}. The resulting scattering matrix $t$ exhibits poles on the
 unphysical sheet which are related to the resonances $\sigma(600)$, $f_{0}(980)$, 
$a_{0}(980)$.

In the case of the subsystem constituted by a heavy and a light pseudoscalar mesons,
 like $DK$, $D\pi$, since the heavy mesons contain both light and heavy quarks, one 
expects both the chiral symmetry of the light quarks and the symmetry of the heavy 
quarks to be considered. Having this in mind, to determine the scattering matrix of 
a heavy meson $H$ with a light pseudoscalar $P$, we follow 
Refs.~\cite{guozou,hanhart}, where the leading order Lagrangian describing this 
interaction is given by the kinetic and mass term of the heavy mesons (chiraly 
coupled to pions), 
\begin{align}
\mathcal{L}=D_{\mu}HD^{\mu}H^{\dagger}-\mathring{M}^{2}_{H} H H^{\dagger}\label{lheavy},
\end{align}
with $H=\left(\begin{array}{ccc}D^{0}&D^{+}&D^{+}_{s}\end{array}\right)$ collecting 
the heavy mesons, whose mass in the chiral limit is $\mathring{M}_{H}$, $P$ is given
 by Eq.~(\ref{phi}) and $D_{\mu}$ is the covariant derivative~\cite{yan}
\begin{align}
D_{\mu}H^{\dagger}&=(\partial_{\mu}+\Gamma_{\mu})H^{\dagger},\nonumber\\
D_{\mu}H&=H(\overleftarrow{\partial}_{\mu}+\Gamma^{\dagger}_{\mu}),\\
\Gamma_{\mu}&=\frac{1}{2}(u^{\dagger}\partial_{\mu}u+u\partial_{\mu}u^{\dagger}),\nonumber\\
u^{2}&=e^{i\sqrt{2}P/f}.\nonumber
\end{align}
For the process which concerns us, i.e., $H P\to HP$, Eq.~(\ref{lheavy}) becomes
\begin{align}
\mathcal{L}_{HP}=\frac{1}{4f^{2}}\left\{\partial^{\mu}H[P,\partial_{\mu}P]H^{\dagger}-H[P,\partial_{\mu}P]\partial^{\mu}H^{\dagger}\right\},
\end{align}
and the lowest order amplitude obtained from this Lagrangian in terms of the 
Mandelstam variables reads as
\begin{align}
V_{ij}=-\frac{C_{ij}}{4f^{2}}(s-u).\label{pot}
\end{align}
In Eq.~(\ref{pot}) $i$ and $j$ represents the initial and final channels, 
respectively, and the $C_{ij}$ coefficients have been earlier calculated and can be 
found in Refs.~\cite{guozou,hanhart}. This potential is further projected on s-wave.

 As in Ref.~\cite{guozou,hanhart}, we consider the coupled channels $DK$, $D_{s}
\eta$ and $D_{s}\pi$ for strangeness $+1$ and $D\pi$, $D\eta$, $D_{s}\bar K$ for 
strangeness $0$. The loop function $\mathcal{G}$ of Eq.~(\ref{BS}) is determined 
using Eq.~(\ref{g}) with $\mu=1000$ MeV and $a=-1.846$~\cite{guo2,hanhart},
obtaining in this way a pole in the $DK$ system for total isospin $0$ at 2318 MeV, 
which corresponds to the state $D_{s^* 0}(2317)$, and a pole at 2446-i43 MeV in the 
$D\pi$ system in isospin 0, associated with the resonance $D^{*}_{0}(2400)$.

Once the two-body scattering matrices are calculated, we can proceed with the 
determination of the three-body $T$ matrix for the $DK\bar K$ system. To do that we 
use the approach of Refs.~\cite{mko1,mko2,mko3,mko4}, in which the Faddeev 
partitions, $T^1$, $T^2$ and $T^3$,  are written as 
\begin{equation}
T^i =t^i\delta^3(\vec{k}^{\,\prime}_i-\vec{k}_i) + \sum_{j\neq i=1}^3T_R^{ij}, \quad i=1,2,3,
\label{Ti}
\end{equation}
with $\vec{k}_{i}$ ($\vec{k}^\prime_{i}$) being the initial (final) momentum of the 
particle $i$ and $t^{i}$ the two-body $t$-matrix which describes the interaction of 
the $(jk)$ pair of the system, $j \neq k\neq i=1,2,3$. The total three-body 
$T$-matrix is obtained by summing the $T^{i}$ partitions,
\begin{equation}
T = T^{1} + T^{2} + T^{3} = \sum_{i=1}^{3}t^i\delta^3(\vec{k}^{\,\prime}_i-\vec{k}_i) +T_{R}\label{T}
\end{equation}
where we define
\begin{equation}
T_{R} \equiv \sum_{i=1}^3\sum_{j\neq i=1}^{3}T^{ij}_{R} . \label{ourfullt}
\end{equation}

The $T^{ij}_{R}$ partitions in Eq.~(\ref{Ti}) satisfy the following set of coupled 
equations

\begin{equation}
T^{\,ij}_R = t^ig^{ij}t^j+t^i\Big[G^{\,iji\,}T^{\,ji}_R+G^{\,ijk\,}T^{\,jk}_R\Big], \quad i\ne j, j\ne k = 1,2,3. 
  \label{Trest}
\end{equation}
where $g^{ij}$ corresponds to the three-body Green's function of the system and its elements are defined as
\begin{eqnarray}
g^{ij} (\vec{k}^\prime_i, \vec{k}_j)=\Bigg(\frac{N_{k}}{2E_k(\vec{k}^\prime_i+\vec{k}_j)}\Bigg)\frac{1}{\sqrt{s}-E_i
(\vec{k}^\prime_i)-E_j(\vec{k}_j)-E_k(\vec{k}^\prime_i+\vec{k}_j)+i\epsilon},\label{Green}
\end{eqnarray}
with $N_{k}=1$ for mesons  and $E_{l}$, $l=1,2,3$, is the energy of the particle 
$l$.

The $G^{ijk}$ matrix in Eq.~(\ref{Trest}) represents  a loop function of 
three-particles and it is written as
\begin{equation}
G^{i\,j\,k}  =\int\frac{d^3 k^{\prime\prime}}{(2\pi)^3}\tilde{g}^{ij} \cdot F^{i\,j \,k}
\label{eq:Gfunc}
\end{equation}
with the elements of  $\tilde{g}^{ij}$ being 
\begin{eqnarray}
\tilde{g}^{ij} (\vec{k}^{\prime \prime}, s_{lm}) = \frac{N_l}
{2E_l(\vec{k}^{\prime\prime})} \frac{N_m}{2E_m(\vec{k}^{\prime\prime})} 
\frac{1}{\sqrt{s_{lm}}-E_l(\vec{k}^{\prime\prime})-E_m(\vec{k}^{\prime\prime})
+i\epsilon}, \quad i \ne l \ne m,
\label{eq:G} 
\end{eqnarray}
and the matrix $F^{i\,j\,k}$, with explicit variable dependence, is given by 
\begin{equation}
F^{i\,j\,k} (\vec{k}^{\prime \prime},\vec{k}^\prime_j, \vec{k}_k,  s^{k^{\prime\prime}}_{ru})= t^{j}(s^{k^{\prime\prime}}_{ru}) g^{jk}(\vec{k}^{\prime\prime}, \vec{k}_k)
\Big[g^{jk}(\vec{k}^\prime_j, \vec{k}_k) \Big]^{-1}
\Big[ t^{j} (s_{ru}) \Big]^{-1},  \quad j\ne r\ne u=1,2,3.\label{offac}
\end{equation}
In Eq. (\ref{eq:G}), $\sqrt{s_{lm}}$ is the invariant mass of the $(lm)$ pair and 
can be calculated in terms of the external variables. The upper index $k^{\prime
\prime}$ for the invariant mass $s^{k^{\prime\prime}}_{ru}$ of Eq.~(\ref{offac}) 
indicates its dependence on the loop variable (see Ref. \cite{mko2} for more 
details).

The $T^{ij}_R$ partitions given in Eq.~(\ref{Trest}) are functions of  the total 
three-body energy, $\sqrt s$, and the 
invariant mass of the particles 2 and 3, $\sqrt{s_{23}}$. The other invariant masses,
 $\sqrt{s_{12}}$ and $\sqrt{s_{31}}$ can be obtained in terms of  $\sqrt s$ and 
$\sqrt{s_{23}}$, as it was shown in Ref. \cite{mko2,mko3}. In this model, peaks obtained in the 
modulus squared  of the three-body $T$-matrix are related to dynamically generated 
resonances. Finally, it should be mentioned that the first term in Eq.~(\ref{T}) can not give rise to any 
state generated due to the three-body dynamics and, hence, we can just study the properties of the $T_R$ matrix 
defined in Eq.~(\ref{ourfullt}).

\begin{figure}[h]
\centering
\includegraphics[width=0.7\textwidth]{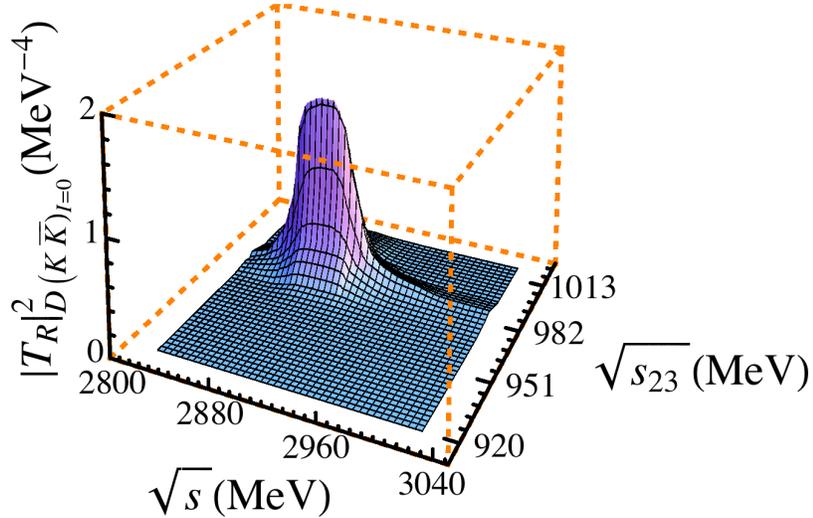}
\caption{Squared amplitude for the $D K\bar K$ channel for total isospin $I=1/2$ with the $K\bar K$ subsystem in isospin zero.}\label{Df0}
\end{figure}

Further, we work in the charge basis, then, to associate the peaks found in the three-body 
$T$-matrix with physical states we need to project $T$ on an isospin basis. We do 
this by defining a basis where the states are labeled in terms of the total isospin 
$I$ of the three-body system and the isospin of one of the two-body subsystems, 
which in the present case is taken as the isospin of the $K\bar K$ subsystem or the 
system made by particles 2 and 3, $I_{23}$, and evaluate the transition amplitude 
$\langle I,I_{23}|T_R|I,I_{23}\rangle$. The isospin  $I_{23}$ can be either 0 or 1, thus, 
the total isospin $I$ can be 1/2,  or 3/2. For the cases involving the states 
$|I=1/2$, $I_{23}=1\rangle$,  $|I=3/2$, $I_{23}=1\rangle$  we find no 
structure which could be related to a resonance or a bound state. Thus, in the 
following, we discuss the case  $I=1/2$ with $I_{23}=0$, where we do find a 
resonance.

In Fig.~\ref{Df0} we show the results obtained for the modulus squared of the 
scattering amplitude of the $DK\bar K$ channel for total isospin $1/2$ with the 
$K\bar K$ system in total isospin 0. A peak around 2890 MeV with a width of 55 MeV 
is found when the $K\bar K$ system is in the isospin zero configuration with an 
invariant mass of around 985 MeV, thus, forming the $f_{0}(980)$ resonance. 

If instead of using the isospin base $|I,I_{23}\rangle$ we use $|I,I_{12}\rangle$, 
with $I_{12}$ the isospin of the (12) subsystem, i.e., the $DK$ system, no clear 
signal for the peak shown in Fig.~\ref{Df0} is observed when the $DK$ system is in 
isospin zero.
 
These results are similar to the ones obtained in the previous section using 
QCD sum rules, in which a state of mass around $\sim 2900$ MeV is found to couple 
more to the $Df_{0}(980)$ current than to the one corresponding to $D_{s^{*}0}(2317)\bar K$.

\section{Summary}
We have studied the $DK\bar K$ system using two different methods: one based on QCD 
sum rules and other on solving few-body equations. In the former case, the $D f_{0}
(980)$ and $D_{s^{*}0}(2317)\bar K$ configurations of the $DK\bar K$ system have 
been investigated and a state with a mass  around 2.9 GeV has been found, which couples more 
to a $D f_{0}$ molecular current.  In the latter, the Faddeev equations 
have been solved with input two-body $t$ matrices which generate the $f_{0}(980)$ 
and $D_{s^{*}0}(2317)$, respectively, in the $K\bar K$ and $D K$ systems and related
 coupled channels. As a result, a state with a mass close to 2.9 GeV and a width 
of 55 MeV was found when the $K\bar K$ subsystem  generates the $f_{0}(980)$ 
resonance. The findings obtained within these two different methods are quite 
similar, hinting towards the existence of a $Df_0(980)$ molecular state with a mass
 close to 2.9 GeV. A state with this mass has not been  discovered experimentally 
so far, and the heaviest known $D$ meson is the $D(2750)$, whose mass is around 150 
MeV below the one  found in this manuscript. We strongly encourage the search of a 
state decaying into $DK\bar{K}$ with the characteristics of the one found here.

\section{Acknowledgements} 
The authors would like to thank the Brazilian funding agencies FAPESP and CNPq for the financial support.


\begin{thebibliography}{200}
 \bibitem{kroko}
  P.~Krokovny {\it et al.}  [Belle Collaboration],
  Phys.\ Rev.\ Lett.\  {\bf 91} (2003) 262002.
  
  \bibitem{aubert} 
 B.~Aubert {\it et al.}  [BABAR Collaboration],
  Phys.\ Rev.\ Lett.\  {\bf 90}, 242001 (2003), {\it ibid} Phys.\ Rev.\ D {\bf 74}, 032007 (2006).
  
  \bibitem{abe} 
  K.~Abe {\it et al.}  [Belle Collaboration],
  Phys.\ Rev.\ D {\bf 69}, 112002 (2004), {\it ibid} Phys.\ Rev.\ Lett.\  {\bf 92}, 012002 (2004).
  
  \bibitem{choi} 
  S.~K.~Choi {\it et al.}  [BELLE Collaboration],
  Phys.\ Rev.\ Lett.\  {\bf 100}, 142001 (2008),
{\it ibid}   
  Phys.\ Rev.\ D {\bf 84}, 052004 (2011).
  
  \bibitem{li} 
 ÊH.~Li,
 Ê
 ÊNucl.\ Phys.\ Proc.\ Suppl.\ Ê{\bf 162}, 312 (2006).
 Ê
  
  \bibitem{mfrev} 
  M.~Nielsen, F.~S.~Navarra and S.~H.~Lee,
  Phys.\ Rept.\  {\bf 497}, 41 (2010).
  

\bibitem{Swanson:2006st}
  E.~S.~Swanson,
  Phys.\ Rept.\  {\bf 429}, 243 (2006).

\bibitem{Zhu:2007wz}
  S.~L.~Zhu,
  Int.\ J.\ Mod.\ Phys.\ E {\bf 17}, 283 (2008).

\bibitem{Brambilla:2010cs}
  N.~Brambilla, {\it et al.},
  Eur.\ Phys.\ J.\  {\bf C71}, 1534 (2011).
 
  
\bibitem{nakahara}
  Y.~Nakahara, M.~Asakawa, T.~Hatsuda,
  Phys.\ Rev.\  {\bf D60} (1999)  091503;

  K.~Sasaki, S.~Sasaki and T.~Hatsuda,
  Phys.\ Lett.\  B {\bf 623} (2005) 208.
  
\bibitem{mathur}
  N.~Mathur, A.~Alexandru, Y.~Chen {\it et al.},
  Phys.\ Rev.\  {\bf D76} (2007) 114505.
 
\bibitem{basak}
  S.~Basak, R.~G.~Edwards, G.~T.~Fleming {\it et al.},
  Phys.\ Rev.\  {\bf D76} (2007) 074504.

\bibitem{bulava}
  J.~Bulava, R.~G.~Edwards, E.~Engelson {\it et al.},
  Phys.\ Rev.\  {\bf D82} (2010) 014507.
  
\bibitem{morningstar}
  C.~Morningstar, A.~Bell, J.~Bulava {\it et al.},
  AIP Conf.\ Proc.\  {\bf 1257} (2010) 779.

\bibitem{svz} M.A. Shifman, A.I. and Vainshtein and V.I. Zakharov,
Nucl. Phys. B {\bf 147}, 385 (1979).

\bibitem{rry} L.J. Reinders, H. Rubinstein and S. Yazaki, Phys. Rept. 
{\bf 127}, 1 (1985).

\bibitem{SNB} For a review and references to original works, see e.g.,
S. Narison, {\it QCD as a theory of hadrons,
Cambridge Monogr. Part. Phys. Nucl. Phys. Cosmol.} {\bf 17}, 1 (2002); {\it QCD spectral sum rules ,  World Sci. Lect. Notes Phys.} 
{\bf 26}, 1 (1989);
{ Acta Phys. Pol.} B {\bf 26}, 687 (1995); { Riv. Nuov. Cim.} {\bf 10N2}, 1
(1987); { Phys. Rept.} {\bf 84}, 263 (1982).
  
\bibitem{colangelo2} 
  P.~Colangelo and A.~Khodjamirian,
  In *Shifman, M. (ed.): At the frontier of particle physics, vol. 3* 1495-1576.
 
 \bibitem{kaiser}
N.~Kaiser, P.B.~Siegel, and W.~Weise, Nucl. Phys. A {\bf 594}, 325 (1995).

\bibitem{osetramos}
  E.~Oset, A.~Ramos,
  Nucl.\ Phys.\  {\bf A635}, 99-120 (1998).

\bibitem{ollerulf}
J. A. Oller and U. G. Meissner, Phys. Lett. B {\bf 500}, 263 (2001).

\bibitem{jido3}
  D.~Jido, A.~Hosaka, J.~C.~Nacher, E.~Oset and A.~Ramos,
  Phys.\ Rev.\  C {\bf 66},  025203 (2002).

\bibitem{jido2}
D. Jido, J. A.Oller, E. Oset, A. Ramos and U. G. Meissner, Nucl. Phys. A {\bf 725}, 181(2003).


\bibitem{oller}
  J.~A.~Oller, E.~Oset,
  Nucl.\ Phys.\  {\bf A620 },  438-456 (1997).

  
 \bibitem{pelaez}
J. A. Oller, E. Oset and J. R. Pelaez, Phys. Rev.  {\bf D59}, 074001 (1999) [Erratum-ibid.
D 60, 099906 (1999 ERRAT,D75,099903.2007)].

 
\bibitem{ollerprog}
  J.~A.~Oller, E.~Oset, A.~Ramos,
  Prog.\ Part.\ Nucl.\ Phys.\  {\bf 45}, 157-242 (2000).
  
  \bibitem{michael1} 
  M.~Doring, U.~-G.~Meissner, E.~Oset and A.~Rusetsky,
  Eur.\ Phys.\ J.\ A {\bf 47}, 139 (2011), {\it ibid} Eur.\ Phys.\ J.\ A {\bf 48}, 114 (2012).
  
  \bibitem{albaladejo} 
  M.~Albaladejo, J.~A.~Oller, E.~Oset, G.~Rios and L.~Roca,
  JHEP {\bf 1208}, 071 (2012).
  
  \bibitem{roca} 
  L.~Roca and E.~Oset,
  Phys.\ Rev.\ D {\bf 85}, 054507 (2012).
  
  \bibitem{martineztorres1} 
  A.~Martinez Torres, L.~R.~Dai, C.~Koren, D.~Jido and E.~Oset,
  Phys.\ Rev.\ D {\bf 85}, 014027 (2012).
  
\bibitem{martineztorres2} 
  A.~Martinez Torres, M.~Bayar, D.~Jido and E.~Oset,
  arXiv:1202.4297 [hep-lat].
  
\bibitem{matheus} 
  R.~D'E.~Matheus, S.~Narison, M.~Nielsen and J.~M.~Richard,
  Phys.\ Rev.\ D {\bf 75}, 014005 (2007).
   
\bibitem{bracco} 
  M.~E.~Bracco, S.~H.~Lee, M.~Nielsen and R.~Rodrigues da Silva,
  Phys.\ Lett.\ B {\bf 671}, 240 (2009)
  
 
\bibitem{narison} S.~Narison, Phys.\ Lett.\ B {\bf 216}, 191 (1989); {\bf 341}, 73 
(1994); {\bf 361}, 121 (1995), {\bf 387}, 162 (1996); {\bf 466}, 345 (1999); 
{\bf 624}, 223 (2005).

\bibitem{narison2} S.~Narison, Phys.\ Lett.\ B {\bf 707}, 259 (2012); R.M.
Albuquerque, X. Liu, M. Nielsen, Phys.\ Lett.\ B {\bf 718}, 492 (2012).
 
\bibitem{io1}
B.~L. Ioffe, Nucl.\ Phys.\ B {\bf 188}, 317 (1981); B {\bf 191}, 591(E) (1981); Prog.\ Part.\ Nucl.\ Phys.\  {\bf 56}, 232 (2006).

\bibitem{width1}   S.~H.~Lee, K.~Morita and M.~Nielsen,
  Phys.\ Rev.\ D {\bf 78}, 076001 (2008).
  
\bibitem{widht2}   H.~G.~Dosch, E.~M.~Ferreira, F.~S.~Navarra and M.~Nielsen,
  Phys.\ Rev.\ D {\bf 65}, 114002 (2002). 

\bibitem{hyodo}
 T.~Hyodo, S.~I.~Nam, D.~Jido, A.~Hosaka,
  Prog.\ Theor.\ Phys.\  {\bf 112}, 73-97 (2004).
  
\bibitem{bruns}
  P.~C.~Bruns, M.~Mai, U.~-G.~Meissner,
  Phys.\ Lett.\  {\bf B697}, 254-259 (2011).
  
   \bibitem{kolomeitsev} 
  E.~E.~Kolomeitsev and M.~F.~M.~Lutz,
  Phys.\ Lett.\ B {\bf 582}, 39 (2004).
  
\bibitem{guozou}
  F.~-K.~Guo, P.~-N.~Shen, H.~-C.~Chiang, R.~-G.~Ping and B.~-S.~Zou,
  Phys.\ Lett.\ B {\bf 641}, 278 (2006).
  
  \bibitem{daniel} 
  D.~Gamermann, E.~Oset, D.~Strottman and M.~J.~Vicente Vacas,
  Phys.\ Rev.\ D {\bf 76}, 074016 (2007).
  
  \bibitem{guo2} 
  F.~-K.~Guo, C.~Hanhart, S.~Krewald and U.~-G.~Meissner,
  Phys.\ Lett.\ B {\bf 666}, 251 (2008)
  
  \bibitem{hanhart} 
  F.~-K.~Guo, C.~Hanhart and U.~-G.~Meissner,
  Eur.\ Phys.\ J.\ A {\bf 40}, 171 (2009).
  
  
  
  \bibitem{gasser} 
  J.~Gasser and H.~Leutwyler,
  Nucl.\ Phys.\ B {\bf 250}, 465 (1985).
  
  \bibitem{meissner} 
  U.~G.~Meissner,
  Rept.\ Prog.\ Phys.\  {\bf 56}, 903 (1993).
  
  \bibitem{ecker} 
  G.~Ecker,
  Prog.\ Part.\ Nucl.\ Phys.\  {\bf 35}, 1 (1995).
  
  \bibitem{burdman} 
  G.~Burdman and J.~F.~Donoghue,
  Phys.\ Lett.\ B {\bf 280}, 287 (1992).
    
  \bibitem{jenkins} 
  E.~E.~Jenkins,
  Nucl.\ Phys.\ B {\bf 412}, 181 (1994) 
    
  \bibitem{yan} 
  T.~-M.~Yan, H.~-Y.~Cheng, C.~-Y.~Cheung, G.~-L.~Lin, Y.~C.~Lin and H.~-L.~Yu,
  Phys.\ Rev.\ D {\bf 46}, 1148 (1992)
  [Erratum-ibid.\ D {\bf 55}, 5851 (1997)].  
    
  \bibitem{mko1}
  A.~Mart\'inez Torres, K.~P.~Khemchandani and E.~Oset,
 Phys.\ Rev.\ C  {\bf 77} 042203 (2008);   
 Eur.\ Phys.\ J.\ A  {\bf 35}  (2008) 295.

  
\bibitem{mko2}
  K.~P.~Khemchandani, A.~Mart\'inez Torres and E.~Oset,
 Eur.\ Phys.\ J.\ A {\bf  37},  (2008) 233.


\bibitem{mko3}
  A.~Mart\'inez Torres et al., 
    Phys.\ Rev.\ D {\bf 78}  (2008) 074031.


\bibitem{mko4}
  A.~Mart\'inez Torres et al., 
 Phys.\ Rev.\ D  {\bf 80},  (2009) 094012.

\bibitem{faddeev} 
  L.~D.~Faddeev,
  Sov.\ Phys.\ JETP {\bf 12}, 1014 (1961)
  [Zh.\ Eksp.\ Teor.\ Fiz.\  {\bf 39}, 1459 (1960)].

 \end{thebibliography}
\end{document}